\documentclass[12pt]{article}
\usepackage{color}
\usepackage{latexsym}
\usepackage{amsmath}
\usepackage{amsfonts}
\usepackage{amssymb}
\usepackage{indentfirst}

\title{Superintegrable  systems on spaces of constant curvature}
\author{Cezary Gonera\thanks{e-mail: cgonera@uni.lodz.pl}, \quad Magdalena Kaszubska\\
\small Department of Theoretical Physics and Computer Science, \\
\small University of \L\'od\'z,\\
\small Pomorska 149/153, 90-236 {\L}\'od\'z, Poland
}
\date{}
\begin{document}
\maketitle 
\begin{abstract}
Construction and classification of 2D superintegrable systems (i.e. systems admitting, in addition to two global integrals of motion guaranteeing the Liouville integrability, the third global and independent one) defined on 2D spaces of constant curvature and separable in the so called geodesic polar coordinates are presented. The method proposed is applicable to any value of curvature including the case of Euclidean plane, the 2-sphere and the hyperbolic plane. The mathematic used is essentially "physical", in particular it refers to the very elegant technique of action-angle variables and perturbation theory so most of our mathematical formulas have a clear physical meaning. The main result can be considered as a kind of generalization of the Bertrand's theorem on 2D spaces of constant curvature
and it covers most of known separable and superintegrable models on such spaces ( in particular, the so-called Tremblay-Turbiner-Winternitz (TTW) and Post-Winternitz (PW) models which have recently attracted some interest).

\end{abstract}
\section{Introduction}
\par The notion of superintegrability seems to be less known than that of integrability. This is in spite of the fact that two simple, yet fundamental (in almost all branches of physics ranging from the atomic physics to the cosmology), systems: the isotropic harmonic oscillator and Kepler problem are superintegrable. Indeed, the rotational symmetry of the Kepler and the isotropic harmonic oscillator potentials implies that the angular momentum is conserved. This can be used to construct, in addition to the energy, two global functionally independent integrals of motion (one can take, for instance, the square of the total angular momentum and one of its components ). All these integrals, when expressed in terms of generalized coordinates and momenta, are in involution (i.e. Poisson commute). A system with three degrees of freedom posesing three global and functionally independent integrals of motion in involution is by the definition integrable in the Liouville sense \cite{b1}. However, it is well known that in the both cases another two additional global and functionally independent integrals of motion can be found. These two extra integrals which do not arise from an explicit geometrical invariance of the potential are constructed out of the famous Runge-Lenz vector \cite {b2} in case of  the Kepler system  and so called Fradkin tensor \cite {b2a} in the oscillator case. The additional  constants together with the Liouville ones generate a higher symmetry. It is $SO(4)$ group (when restricted to the sub-manifold of constant energy) in the case of Kepler problem  \cite {b3} and $SU(3)$ group in the isotropic harmonic oscillator case  \cite {b4}.
\par The integrable systems admitting more (global and functionally independent ) first integrals than degrees of freedom are called superintegrable  \cite {b5}. If they have  maximal possible number of independent constants, i.e. $2n - 1$ (for the system with n degree of freedom) they are called maximally superintegrable. The Kepler model and isotropic harmonic oscillator provide the canonical examples of such systems.
\par If n-dimensional sub-manifold of phase space determined by the n involutive first integrals is compact and connected it is topologically equivalent to the n-dimensional so called Arnold - Liouville torus ( in general, it is the product of a torus and Euclidean space)  \cite {b1}. Now, due to the existence of additional constants of motion the trajectories of superintegrable systems are restricted to lower dimensional sub-manifolds of Arnold-Liouville tori. In the particular case of maximally superintegrable systems, when the number of global independent integrals o motion increases to $2n - 1$ the classical trajectories are closed curves. It is so, of course, in the Kepler and isotropic harmonic oscillator cases. Actually, the fact that all trajectories of bounded motions are closed distinguishes and characterize these two simple systems in the unique way. This is due to the old and very elegant Bertrand's theorem  \cite {b6} which states that the only central potentials for which all bounded trajectories are closed are just Kepler and isotropic oscillator ones. From the superintegrability point of view the Bertrand theorem provides a complete classification of 3-D superinegrable systems with central potentials. In fact, due to the rotational symmetry of these potentials Bertrand's result provides also a complete classification of two-dimensional isotropic superintegrable systems.
\par In general, a dynamics in non-central potentials is much more complicated than in the central ones. Consequently, a search for superintegrable systems in non-central fields is more involved and one can hardly expect that a non-central counterpart of Bertrand's theorem exists. Nevertheless, there is a number of papers devoted to the study of the superintegrability in non-central potentials, both in the Euclidean and curved configuration spaces  \cite {b7} -  \cite {b12}. In particular, the so-called Tremblay-Turbiner-Winternitz (TTW) \cite {b13},  \cite {b14} and Post-Winternitz (PW)  \cite {b15} models which have recently attracted some interest (see for example Refs. \cite {b16} -  \cite {b23} and references therein ) provide examples of superintegrable systems with non-central potentials defined on Euclidean plane while spherical and pseudospherical  generalizations of these models ( considered in  \cite {b24}) represent non-isotropic superintegrable systems on curved configuration spaces.
\par Actually, both TTW and PW classical models on Euclidean plane belong, respectively, to one of two families of superintegrable systems found many years ago by Onofri and Pauri  \cite {b25}. These authors managed to classify all superintegrable systems defined on 2D Euclidean plane with Hamiltonians separable in the polar coordinates. It appears that all such systems can be divided into two classes called by the authors, respectively, Kepler and oscillator families (the former one contains in particular the Kepler model while the latter, among others, the isotropic harmonic oscillator). Unfortunately, it seems that this very nice and interesting result is not as well known as it deserves to be. This is perhaps due to a rather involved method of derivation the authors used.
\par In the present paper we demonstrate a construction and classification of 2D superintegrable systems defined on 2D spaces of constant curvature and separable in so called geodesic polar coordinates. Our method works for any value of the curvature including the case of Euclidean plane, the 2-sphere and the hyperbolic plane. The main results, which agree  with those of Onofri and Pauri in the Euclidean plane case, can be considered as a further generalization of Bertrand's theorem on the 2D spaces of constant curvature. 
\par The paper is organized as follows. In section two we set our notation and, using the so called geodesic polar coordinates, we write out the metric of constant curvature $k$ describing in a uniform way a geometry of 2D sphere, 2D Euclidean and 2D hyperbolic plane. Then we introduce the general Hamiltonian separable in the geodesic polar coordinates and explain our main task, that is a construction and classification of radial and angular potentials leading to superintegrable dynamics.
\par A necessary and sufficient condition for integrable system to be superintegrable is recalled in section three. In the framework of elegant technique of action-angle variables it states that Hamiltonian of superintegrable system has to be a function of linear combination of action variables with integer coefficients. Then we show that this condition, when applied to 2D integrable and separable system, can be formulated in the form of an equality (up to an integer factor) of radial and angular periods of motion. The radial period corresponds to the dynamics in an isochronous (i.e. such that the period of motion does not depend on energy) effective potential being determined by the radial potential entering the original Hamiltonian. This simple consequence of the superintegrability condition plays a key role in our method. First, it implies that search for our superintegrable systems can be started with the construction of the effective isochronous radial potentials (actually its $ \tilde V_\sigma (\rho )$ part being directly related to the radial potential $V_\sigma (r)$ entering the original Hamiltonian). This is done in the section four, where the relevant equation for $\tilde V_\sigma (\rho )$ has been introduced and solved. Knowing $\tilde V_\sigma (\rho )$ potentials allows us to find the original radial potentials $V_\sigma (r )$ leading to superintegrable systems. Next, having these potentials we refer to our key condition and calculate the periods of angular motions. Finally, considering the formula for a period of one-dimensional motion in a potential as the integral equation (actually it is the equation of Abel's type) with this potential being unknown function we find the angular potentials corresponding to the periods of angular motions.\\
Section four contains also the discussion of the explicit forms of the corresponding radial action variables. It is explained how these actions and superintegrability condition can be used to find the explicit forms of angular action variables as well as to determine, in the independent way, the relevant periods of motions. At the end of this section it is demonstrated how superintegrable models which have recently attracted some attention are accommodated in the general families constructed in the paper. We also briefly indicate that in the central potential case our method reproduces, as it should be, the famous Bertrand result for Euclidean plane and its generalizations for curved spaces  \cite {b26} -  \cite {b33}.
\par The summary of our approach and short discussion of the results can be found in the last section four.

\section{Classification of 2D-superintegrable systems separable in the "geodesic polar" coordinates}
\par
It is known that on any two-dimensional Riemannian space one can introduce, at least locally, the so called "geodesic polar" coordinates. These
coordinates, based on an origin point $O$ and the oriented geodesic $l_0$ through $O$ are defined as follows. For any point $P$ in some suitable
neighborhood of $O$ one takes the unique geodesic $l$ joining $O$ and $P$. Then the (positive) distance $r$ between $O$ and $P$ measured along $l$, and 
the angle $\varphi $ between $l$ and $l_1$ measured around $O$ define the geodesic polar coordinates $(r,\varphi )$ of $P$. Obviously, these
coordinates reduce to the usual polar ones in the Euclidean plane case.\\
The positive-defined metric of constant curvature $k$ describing the geometry of 2D sphere (corresponding to $k > 0$), 2D Euclidean plane (corresponding
to $k = 0$) and 2D hyperbolic plane (corresponding to $k < 0$), when written in terms of geodesic polar coordinates $(r, \varphi ),$ reads

\begin{equation}
\label{e1}
ds^2 = dr^2 + s_k^2(r)d\varphi ^2,
\end{equation}
\;\\\
where $s_k(r)$ functions are defined as follows.

\begin{equation}
\label{e2}
s_k(r) = 
\begin{cases}
\frac{1}{\sqrt{k}} sin(\sqrt{k}r) & k > 0\\
 r & k = 0\\
 \frac{1}{\sqrt{-k}} sh(\sqrt{-k}r) & k < 0\\
\end{cases}
\end{equation}
\;\\\
Consequently, the general natural Lagrangians ( kinetic energy minus potential one)  and the corresponding Hamiltonians take the forms

\begin{equation}
\label{e3}
L_\sigma (r,\varphi , \dot r ,\dot \varphi ) = \frac{1}{2}( \dot r^2 + s_k^2(r)\dot \varphi^2 )  - U_\sigma ( r, \,\varphi)
\end{equation}

\begin{equation}
\label{e4}
H_\sigma (r,\varphi , p_{r}, p_{\varphi }) = \frac{p_{r}^{2}}{2} + \frac{p_{\varphi}^{2}}{2s_k^2(r)} + U_\sigma ( r, \,\varphi) 
\end{equation}
\;\\\
where $U_\sigma ( r, \,\varphi) $ represents potential energy, $p_{r}$ and $p_{\varphi}$ denote canonical momenta conjugated to 
$r,\varphi$ coordinates respectively, while $\sigma  = sign k$.\\
Hence the Hamiltonian separable in the polar coordinates reads
\;\\\
\begin{equation}
\label{e5a}
H(r,\varphi , p_{r}, p_{\varphi }) = \frac{p_{r}^{2}}{2} + \frac{1}{2s_k^2(r)}(\frac{p_{\varphi}^{2}}{2} + U_\sigma (\varphi))+ V_\sigma (r)
\end{equation}
\;\\\
In the above equation $U_\sigma (\varphi))$ and $V_\sigma (r)$ denote angular and radial potentials, respectively.

The problem of classification of superintegrable systems with Hamiltonians separable in the polar coordinates can be put in the most direct and simple way as follows. For the Hamiltonians given by eq. (\ref{e5a}) find all radial $V_\sigma (r)$ and angular $U_\sigma (\varphi)$ potentials implying superintegrability that is, in our case, the existence of three functionally independent, globally defined integrals of motion. 
Actually, from the superitegrability point of view, it is natural to assume from the very beginning that our Hamiltonians are Liouville integrable with separation constants: the total energy E and "generalized momentum" A being globally defined integrals. They define a compact and sufficiently regular surface isomorphic to the Liouville - Arnold tori. Then the superintegrability of our systems amounts to the existence of a third independent and global integral.

In the Euclidean plane (i.e. the zero curvature case ), the problem of classification of the radial and angular potentials admitting such third integral was solved by Onofri and Pauri \cite{b25}. In the very nice paper they managed to show that the allowed radial potential should be of the following form  
\begin{equation}
\label{e6}
\begin{split}
V(r) = \frac{C_{1}}{r^{2}} + C_{2}r^{2}\qquad oscillator family \\
\;\\\
V(r) = \frac{C_{3}}{r^{2}} + \frac{\sqrt{C_{4}r^{2} + C_{5}}}{r^{2}} \qquad  "Kepler" family
\end{split} 
\end{equation}
\;\\\
where $C_1, C_2, ...C_5 $ are some constants chosen in such a way that these potentials are physically reasonable, that is they are real and prevent the phenomenon of the falling on a center. All relevant angular potentials are determined up to a function by solving the integral equation of Abel type. To arrive at these results Onofri and Pauri proceeded in the spirit of Landau approach \cite{b34} and transformed periodicity (superintegrability ) condition into the system of integral equations of Abel type with potentials entering as unknown functions. It appears that direct application of this method to nonzero curvature case results in very complicated nonlinear differential equations.\\ 
Our approach is different. Roughly speaking, we use the superintegrability condition to write out the differential equation for the radial potentials and then integral equation to determine the angular one.

\section{Superintegrability condition and its consequences}
\label{s3}

The superintegrability condition and its consequences to be presented below provide a convenient starting point to classify our superintegrable systems\\
As already mentioned, from the superintegrability point of view one can assume from the very beginning, without loosing generality that:\\
 
i) our separable Hamiltonian (see eq.(\ref{e5a})) $H_\sigma (r,\varphi , p_{r}, p_{\varphi })$ and "generalized momentum" \\

\begin{equation}
\label {e7}
L_\sigma ( \varphi ,p_{\varphi }) = \frac{p_{\varphi}^{2}}{2} + U_\sigma (\varphi) 
\end{equation}
\;\\
provide globally defined (and, obviously, involutive) integrals of motion implying Liouville integrability of the system.\\

ii) equations :\\
\begin{equation}
\label{e8}
\begin{split}
 H_\sigma (r,\varphi , p_{r}, p_{\varphi }) = E\\ 
 L_\sigma ( \varphi ,p_{\varphi }) = A 
 \end{split}
\end{equation}
\;\\\
define, for some intervals of the values of separation constants E and A  a compact and sufficiently regular surface ( which is isomorphic, by Arnold-Liouville theorem, to 2-d tori).\\
Then, 
proceeding in the spirit of the elegant classical approach one can introduce the action- angle variables ($\Psi _{\sigma r}, J_{\sigma r}, \Psi _{\sigma \varphi }, J_{\sigma \varphi }$).\\
In our case the relevant action variables are defined by following equations

\begin{equation}
\label{e8a}
\begin{split}
J_{\sigma \varphi }(A) = \frac{1}{\pi }\int_{\varphi _{min}}^{\varphi _{max}}\sqrt{2(A - U_\sigma (\varphi ))}d\varphi\\
\;\\\ 
J_{\sigma r}(A, E) = \frac{1}{\pi }\int_{r_ {min}}^{r _{max}}\sqrt{2(E- V_\sigma (r) - \frac{A}{s_k^{2}(r)})}dr 
\end{split}
\end{equation}
\;\\\\
where $\varphi _{min}$, $\varphi _{max}$, $r_ {min}$, $r _{max}$ are the roots of the relevant integrands.  

 As it is well known in the integrable case the Hamiltonian depends on action variables only i.e. 
 
\begin{equation}
\label{e9}
 H_\sigma  = H_\sigma (J_{\sigma r} , J_{\sigma \varphi} ),
\end{equation}
\;\\\
hence the canonical Hamiltonian equations read 

\begin{equation}
\label{e10}
\begin{split}
\dot \Psi_{\sigma r} = \frac{\partial{H_\sigma }}{\partial{J_{\sigma r}}} \equiv \omega _{r} \qquad  \dot{J}_{\sigma r} = 0 \nonumber \\
\dot \Psi_{\sigma \varphi } = \frac{\partial{H_\sigma }}{\partial{J_{\varphi }}} \equiv \omega _{\varphi } \qquad  \dot{J}_{\varphi \varphi } = 0
\end{split} 
\end{equation}

It is known (see \cite{b34}, for instance) that the existence of the third, independent and globally defined integral is equivalent to the following condition
\;\\\ 
\begin{equation}
\label{11}
H= H(J_{\sigma r} + qJ_{\sigma \varphi} ) \qquad q = \frac{m}{n}, \qquad m,n \in Z 
\end{equation}
\;\\\
Indeed, then (and only then) one can define the third integral of motion, taking for example

\begin{equation}
\label{e12}
 Y \equiv  f(J_{\sigma r} , J_{\sigma \varphi} )\sin(m\Psi_{r} - n\Psi _{\varphi })
\end{equation}
\;\\\
where f is any differentiable functions of action variables.\\
It follows immediately  from equations of motion that $\dot Y = 0 $.\\
\par So, one concludes that the equation
\;\\ 
\begin{equation}
\label{13}
E= H(J_{\sigma r} + qJ_{\sigma \varphi} ) \qquad q = \frac{m}{n}, \qquad m,n \in Z 
\end{equation}
\;\\\
which equivalently can be rewritten as
\;\\\
\begin{equation}
\label{e14}
\zeta (E) =J_{\sigma r}(A, E) + qJ_{\sigma \varphi}(A) 
\end{equation}
\;\\\
where $\zeta$ denotes a smooth function of E, 
provides the necessary  and  (sufficient if q is rational) condition of superintegrability of our systems.\\ 
\par Taking the derivative of eq.(\ref{e14}) with respect to A parameter gives
\;\\\ 
\begin{equation}
\label{e15}
\frac{\partial J_{\sigma r}(A, E)}{\partial A} = - q\frac{\partial J_{\sigma \varphi}(A)}{\partial A} 
\end{equation}
\;\\\
which, in turn implies that the derivative of  radial action $J_{\sigma r}$ with respect to A  i.e.\\
\;\\\ 
\begin{equation}
\frac{\partial J_{\sigma r}(A, E)}{\partial A} \nonumber
\end{equation}
\;\\\
 does not depend on energy E! (as R.H.S. of eq.(\ref{e15}) does not depend on energy).

\par Now, taking into account the definitions of action variables (see eqs.(\ref{e8a})) one finds that the derivatives of these actions with respect to A have the following forms
\;\\\
\;\\\
\begin{equation}
\label{e16}
\begin{split}
\frac{\partial J_{\sigma \varphi }}{\partial A} = \frac{1}{\pi }\int_{\varphi _{min}}^{\varphi _{max}}
\frac{d\varphi }{\sqrt{2(A - U_\sigma (\varphi ))}} \equiv \frac{T_{\sigma \varphi }(A)}{2\pi } \equiv \frac{1}{\Omega _{\varphi }} \\
\;\\\
\;\\\
\frac{\partial J_{\sigma r }}{\partial A} = \frac{-1}{\pi }\int_{\rho  _{min}}^{\rho  _{max}}
\frac{d\rho  }{\sqrt{2(E -  W_{\sigma eff}(\rho  ))}} \equiv \frac{T_{\sigma \rho  }(A)}{2\pi } \equiv \frac{1}{\Omega _{\rho  }}
\end{split} 
\end{equation}
\;\\\
where after the appropriate changes of variables i.e.
\;\\
\begin{equation}
\label{e17}
\begin{split}
r = \frac{1}{\sqrt{k}}arcctg\frac{\rho }{\sqrt{k}} \qquad  k > 0 \\
r=\frac{1}{\rho }\qquad  k = 0 \\
r = \frac{1}{\sqrt{-k}}arccth\frac{\rho }{\sqrt{-k}} \qquad  k < 0  
\end{split}
\end{equation}
\;\\
an effective potential energy $W_{\sigma eff}(\rho  ))$ reads:

\begin{equation}
\label{e18}
 W_{\sigma eff}(\rho ) \equiv  \tilde V_\sigma (\rho ) + A\rho ^2 \pm Ak^2
\end{equation}
\;\\\
 In the above equation the plus sign corresponds to the positive curvature k while the minus to negative one and $\tilde V_\sigma (\rho )$ functions are related to the original  radial potentials  $ V_\sigma (\rho )$ in the following way\\
\newpage
a) for $ \qquad k < 0 \qquad ( \sigma  = -1)$\\
\;\\
\begin{equation}
\label{e20}
\tilde V_{-}(\rho) = V_{-}(\frac{1}{\sqrt{-k}}arccth(\frac{\rho }{\sqrt{-k}})) \qquad i.e. \qquad V_{-}(\rho) = \tilde V_{-}(\sqrt{-k}cth(\sqrt{-k}\rho ))\\
\end{equation}
\;\\\
b) for $ \qquad k = 0\qquad  (\sigma  = 0)$
\;\\\
\begin{equation}
\label{e21}
\tilde V_{0}(\rho) = V_{0}(\frac{1}{\rho }) \qquad i.e. \qquad V_{0}(\rho) = \tilde V_{0}(\frac{1}{\rho} )\\
\end{equation}
\;\\\
c) for $ \qquad k > 0 \qquad (\sigma  = 1)$
\;\\\
\begin{equation}
\label{e22}
\tilde V_{+}(\rho) = V_{+}(\frac{1}{\sqrt{k}}arcctg(\frac{\rho }{\sqrt{k}})) \qquad i.e.\qquad V_{+}(\rho) = \tilde V_{+}(\sqrt{k}ctg(\sqrt{k}\rho) )
\end{equation}
\;\\\

Equations ( \ref{e16}) are  nothing but the periods (divided by $2\pi $ factor ) of one-dimensional motions in the angular $ U_\sigma (\varphi )$ and the effective radial  $W_{\sigma eff}(\rho )$ potentials, respectively.
So, the main implication of the superintegrability condition ( given by eq.(\ref{e15})) is expressed as:

\begin{equation}
\label{e23}
T_{\sigma \rho  }(A) = qT_{\sigma \varphi   }(A) 
\end{equation}
\;\\\
with $T_{\sigma \varphi   }(A)$ and  $T_{\sigma \rho  }(A)$ defined by eqs. (\ref{e16}) and the $T_{\sigma \rho  }(A)$ period NOT depending on
energy E.\\  
Now, due to this energy independence of $T_{\sigma \rho  }(A)$ as well as due to the eqs.(\ref{e20}) - (\ref{e22}) relating radial
 $\tilde V_\sigma (\rho )$ potentials to the original $ V_\sigma (\rho )$ ones the construction and classification of superintegrable systems with Hamiltonians separable in the geodesic polar coordinates can be reduced to the following steps.\\
First, one finds $\tilde V_\sigma (\rho )$ potentials leading to isochronous effective radial potentials $W_{\sigma eff}(\rho )$ i.e. potentials generating motions with energy independent periods.\\  
In the second, intermediate step, knowing the explicit form of $\tilde V_\sigma (\rho )$ or $V_\sigma (\rho )$ one can calculate the periods
of the corresponding angular motions. Indeed, taking into account the second of eqs.(\ref{e16}) one can rewrite the eq.(\ref{e23}) in the form: 

\begin{equation}
\label{e24}
T_{\sigma \varphi   }(A) = \frac{1}{q}T_{\sigma \rho }(A)= -\frac{2\pi }{q} \frac{\partial J_{\sigma r }}{\partial A}     
\end{equation}
\;\\\
So, the periods $T_{\sigma \varphi   }(A)$ of angular motions can be calculated by performing directly integrals given by the second of eqs.(\ref{e16})
or by finding explicit forms   of the radial actions $J_{\sigma r}(A,E)$ first and then taking its derivative with respect to A parameter.\\
Finally, knowing the period of $T_{\sigma \varphi   }(A)$ of angular motion as a function of A ( which plays here the role of energy in 1-d motion in
the angular potential $U_\sigma (\varphi )$ ) one can determine this potential ( up to an arbitrary function). Roughly speaking, this is due to the possibility of regarding the expression for the period of one-dimensional motion in a potential as the integral equation (actually, it is the integral equations  of Abel type) with the potential entering as unknown function.

\section{The construction of potentials defining superintegrable Hamiltonians}
\label{s5}

As explained in the previous section the first step of our method of searching for superintegrable separable Hamiltonians consists in constructing  isochronous effective radial potentials given by eq. (\ref {e18}), i.e.

\begin{eqnarray}
\nonumber
 W_{\sigma eff}(\rho ) = \tilde V_\sigma (\rho ) + A\rho ^2  \pm k^2A \
\end{eqnarray}

To this end we introduce and solve differential equation  on $\tilde V_\sigma (\rho )$ functions. We start with an assumption 
that the effective potential $ W_{\sigma eff}$ attains a local minimum at some $\rho = \rho _{0} \equiv \rho _{0}(A)$ 

\begin{equation}
\label{e25}
 E_{0} = W_{\sigma eff}(\rho_{0} ) =  \tilde V_\sigma (\rho_{0} ) + A {\rho_{0}} ^2  \pm k^2A
\end{equation}

Vanishing of the first derivative of  the effective potential $\ W'_{\sigma eff}(\rho_{0} ) = 0$ at $\varrho _0$ implies that

\begin{equation}
\label{e26}
  A = -\frac{\tilde V'_\sigma (\rho_{0}(A))}{2\rho_{0}(A)}  
\end{equation}

and Taylor expansion of the effective potentials around $\rho _{0}$ reads

\begin{equation}
\label{e27}
 W_{\sigma eff}(\rho ) = E_{0} + \frac{1}{2}{\omega _{0}}^2 (\rho  - \rho _{0})^2 + 
\frac{1}{3}\alpha (\rho  - \rho _{0})^3 + \frac{1}{4}\beta  (\rho  - \rho _{0})^4 
\end{equation}

where

\begin{equation}
\label{e28}
\begin{split}
 \omega ^{2}_{0} =  W''_{\sigma eff}(\rho_{0} ) = \tilde V''_\sigma (\rho_{0}) -  \frac{\tilde V'_\sigma (\rho_{0})}{\rho_{0}} \\
 \;\\\
 \alpha =  \frac{1}{2} W'''_{\sigma eff}(\rho_{0}) = \frac{1}{2} \tilde V'''_\sigma (\rho_{0})\\
 \;\\\
 \beta =  \frac{1}{6} W^{(IV)}_{eff}(\rho_{0}) = \frac{1}{6} \tilde V_{\sigma }^{(IV)}(\rho_{0})
\end{split} 
\end{equation}

Requiring the frequency $\Omega _{\rho }$ of motion in the effective potential $ W_{\sigma eff}$ to be energy independent implies that when one calculates this frequency perturbatively  one obtains at every order the same (and in fact, the exact) result.\\   
In particular, in  harmonic approximation when only terms quadratic in $\rho $ are taken into account one gets
\begin{equation}
\label{e29}
 \Omega _{\rho } = \omega _{0}
\end{equation}

On the other hand, in anharmonic  approximation when terms up to fourth order are kept one has (see, for instance, \cite{b34})   
\begin{equation}
\label{e30}
 \Omega _{\rho } = \omega _{0} + (\frac{3}{4}\beta - \frac{5}{6}\frac{\alpha ^2}{ \omega ^{2}_{0}})\frac{a^2}{2\omega _{0}}  
\end{equation}
with a denoting an amplitude of small oscillations (in zeroth order $E-E_{0}= \frac{1}{2}\omega ^{2}_{0}a$ )\\
Now, as explained earlier, our effective potential is to be isochronous. It means that eq.(\ref{e29}) gives the exact (energy independent) frequency:  $\Omega _{\rho } = \omega _{0}$. Consequently, it follows from eq.(\ref{e30}) that

\begin{equation}
\label{e31}
\frac{3}{4}\beta - \frac{5}{6}\frac{\alpha ^2}{ \omega ^{2}_{0}} = 0  
\end{equation}

In terms of $\tilde V_\sigma $ derivatives it reads

\begin{equation}
\label{e32}
 3\tilde V_\sigma ^{(IV)}(\rho_{0}(A)) =5\frac{\tilde V_\sigma '''(\rho_{0}(A))} {\tilde V_\sigma ''(\rho_{0}(A)) -  \frac{\tilde V_\sigma '(\rho_{0}(A))}{\rho_{0}(A)}}  
\end{equation}

This condition  should hold for any value of A i.e. $\rho _{0}(A)$. So eq.(\ref{e32}) can be regarded as the differential equation

\begin{equation}
\label{e33}
 3\tilde V_\sigma ^{(IV)}=5\frac{\tilde V_\sigma '''} {\tilde V_\sigma '' -  \frac{\tilde V_\sigma '}{\rho}}  
\end{equation}
\;\\\
for $\tilde V_\sigma (\rho )$ potential providing isochronous  $ W_{\sigma eff}(\rho )$ effective one.\\
In spite of unfriendly appearance, eq.(\ref{e33}) can be solved. A series of substitutions:
$\tilde V_\sigma ' = f$,   $g = f' - \frac{f}{\rho }$,   $w = \frac{g}{\rho g'}$  leads to simple differential equations
 
\begin{equation}
\label{e34}
\begin{split}
 -4g = \rho g' \\
 \;\\\
 or \qquad -3\rho w' = 8w^2 + 10w +2
\end{split}  
\end{equation}
\;\\\
which provide respectively, two families of solutions\\
\;\\\
\begin{equation}
\label{e35}
\begin{split}
 \tilde V_\sigma (\rho ) = \gamma \rho ^2 +\frac{\delta }{\rho ^2} \\
 \;\\\
  or \qquad \tilde V_\sigma (\rho ) = B\rho ^2 - \rho \sqrt{D + F\rho ^2}
\end{split} 
\end{equation}
\;\\
where $\delta ,\gamma , B, D, F $ are integration constants.\\
\par Now, taking into account the eqs.(\ref{e20}), (\ref{e21}) and (\ref{e22}) which relate the auxiliary potentials $\tilde V_\sigma $ to the original 
$ V_\sigma$ ones (entering our initial Hamiltonians ( see eq.(\ref{e5a}))) one finds  the explicit forms of radial potentials $ V_\sigma(r)$ relevant for a given curvature sign and necessary for superintegrability. This completes the first step of our construction. It appears that for each curvature sign there exist two families of relevant potentials. In particular, in the zero curvature case these families are nothing but the ones found by Onofri and Pauri \cite{b25}. Following these authors, one of these families will be named oscillator while the other the Kepler one. The explicit form of the potentials corresponding to the different signs of the curvature k are presented below.\\

{\bf a) the negative curvature k case}\\

\par i) the oscillator family\\

\begin{equation}
\label{e36}
 V_-(r) = \frac{\gamma \mid k\mid }{\tanh^2({\sqrt{-k}r})} + \frac{\delta }{\mid k\mid }\tanh^2({\sqrt{-k}r}) 
\end{equation}
\;\\\

\par ii) the Kepler family\\
 
\begin{equation}
 \label{e37}
 V_-(r) = \frac{B\mid k\mid }{\tanh^2{(\sqrt{-k}r)}} - \frac{\mid k\mid }{\tanh^2{(\sqrt{-k}r)}}\sqrt{\frac{D}{\mid k\mid }\tanh^2{(\sqrt{-k}r)} + F}
\end{equation}
\;\\\
\;\\\ 
{\bf b) the zero curvature k case}\\

i) the oscillator family\\

\begin{equation}
\label{e38}
 V_0(r) = \frac{\gamma}{r^2} + \delta r^2 
\end{equation}

ii) the Kepler family\\
 
\begin{equation}
 \label{e39}
 V_0(r) = \frac{B}{r^2} - \frac{\sqrt{Dr^2 + F}}{r^2}
\end{equation}
\;\\\
\;\\\
{\bf c) the positive curvature k case}\\

i) the oscillator family\\

\begin{equation}
\label{e40}
 V_+(r) = \frac{\gamma k }{\tan^2{(\sqrt{k}r)}} + \frac{\delta }{ k}\tan^2{(\sqrt{k} r)} 
\end{equation}

ii) the Kepler family\\

\begin{equation}
 \label{e41}
 V_+(r) = \frac{B k}{\tan^2{(\sqrt{k}r)}} - \frac{ k }{\tan^2{(\sqrt{k}r)}}\sqrt{\frac{D}{k}\tan^2{(\sqrt{k}r)} + F}
\end{equation} 
\;\\\
 To have physically interesting potentials we choose $\gamma ,\delta , B, D, F \geq 0$ and $ B+ A +\sqrt{F} \geq 0 $ as well as $\gamma  + A\geq 0$ \\
Note that  

\begin{equation}
\label{e42}
\begin{split}
 V_-(r) = V_0(\frac{1}{\mid k\mid }\tanh{(\sqrt{-k}r)}) \\
 V_+(r) = V_0(\frac{1}{k}\tan{(\sqrt{k}r)}
\end{split} 
\end{equation}

Knowing the above radial potentials one can pass to the construction of compatible ( from the superintegrability point of view) angular ones. The key condition  here, following from the assumed superintegrability (see eq.\ref{e14}) is given by eq.(\ref{e24}). It says that the period of one-dimensional motion in the angular potential $U_\sigma (\varphi )$ is given  by the derivative  of radial action variable with respect to the A parameter playing here a role of "total energy" . These actions or directly  their derivatives can be computed once the explicit forms of radial potentials are given. Then, having the period $T_{\sigma \varphi }(A)$ of one-dimensional motion in an angular potentials 
$U_\sigma (\varphi )$ as a function of "energy" A one can determine this potential (by solving the integral equation of Abel type with this potential entering as unknown function) \cite{b34}.\\
So, following a shorter and more direct path one calculates the periods $T_{\sigma \varphi }(A)$ of the angular motions by performing integrals given by the second eq.(\ref{e16}). The effective potential entering  the RHS is defined by eq.(\ref{e18}) while $\tilde V(\rho )$ potentials are given by eq.(\ref{e35}). As expected, the results do not depend on the curvature k. This is basically due to the fact that the curvature k enters the effective potentials only through the terms that can be absorbed into the energy E (which, in turn, does not affect the periods of motion in the effective isochronous potentials). So, we have the following formulas for the periods of angular motions as explicit functions of A parameter:
\;\\\
\begin{equation}
\label{e43}
T_{\varphi }(A) = \frac{1}{q\sqrt{2}}\frac{1}{\sqrt{A + \gamma }} 
\end{equation}
 \;\\\
in oscillator family case, and
\;\\\
\begin{equation}
 \label{e44} 
T_{\varphi } (A)= \frac{1}{q\sqrt{2}} ( \frac{1}{\sqrt{A+B + \sqrt{F}}} + \frac{1}{\sqrt{A+B - \sqrt{F}}} )
\end{equation}
\;\\\
\;\\\
for "Kepler" family. \\
\par The above results can be checked in independent way by explicit computation of the radial action variables defined by eq. (\ref{e8a}) with the radial potentials  $V_\sigma (r)$ given by eqs.(\ref{e36}) - (\ref{e41}) and then taking the derivative with respect to A parameter. It appears that in the zero curvature limit the radial actions corresponding to both positive and negative curvature signs tend to the action corresponding to the flat (i.e. $k = 0$) case. 
Now, if we know the explicit forms of the radial actions $J_{\sigma r}(A,E)$, the eq.(\ref{e14}), $J_{\sigma r}(A,E) =\zeta (E) - qJ_{\sigma \varphi }(A)$, allows us to determine (up to a rational multiplicative constant q) the angular actions variables $J_{\sigma \varphi }(A)$. The only poit is that the separation into $\zeta (E)$ and $qJ_{\sigma r}(A,E)$ is defined by eq.(\ref{e14}) up to a constant (independent of A and E); however, this constant can be chosen at will by imposing a normalization on angular potential $U_\sigma (\varphi )$. For instance, one can assume that  $U_\sigma (\varphi )$ attains minimum at  $U_\sigma (\varphi ) = 0$; then $J_{\sigma \varphi }(0) = 0$ which povides the additional condition yielding the separation unique.
A direct inspection of the explicit formulas for the radial actions $J_{\sigma r}(A,E)$ shows that the curvature k enters only their energy dependent parts. Hence, the angular action variables take the same form for all three signs of the curvature and they read

\begin{equation}
\label{e45}
qJ_{\sigma \varphi }(A) = \frac{1}{\sqrt{2}}\sqrt{A + \gamma } \qquad \sigma = 0, \pm1
\end{equation}
\;\\\
\;\\\ 
in oscillator family case, and
\;\\\
\;\\\
\;\\\
\begin{equation}
 \label{e46} 
qJ_{\sigma \varphi }(A) = \frac{1}{\sqrt{2}}(\sqrt{A+B + \sqrt{F}} + \sqrt{A+B - \sqrt{F}}\quad ) \qquad \sigma = 0, \pm1
\end{equation}
\;\\\
\;\\\
\;\\\
for "Kepler" family. \\
\par It follows from the general expression for  $J_{\sigma \varphi }(A)$ (see  eq.(\ref{e8a})) that in the large $A\gg 1$ limit
$qJ_{\sigma \varphi }(A) = \frac{q}{\pi }\sqrt{2A }\Delta \varphi$, where $\Delta \varphi$ corresponds to the the length of the domain of the angular potential $U_\sigma (\varphi )$ (for example, for the TTW model, see eq.(\ref{e56}), $\Delta \varphi = \frac{2\pi}{2n}$). Comparing that with 
large $A\gg 1$ limit of explicit formulas for $qJ_{\sigma \varphi }(A)$ (see eqs.(\ref{e45}) and (\ref{e46})) implies that $ \Delta \varphi = \frac{\pi }{2q}$ in the oscillator case, and $\Delta \varphi =\frac{\pi }{q}$ in the "Kepler" one. In both cases, the q parameter controls the range of angular polar coordinates. \\

Leaving aside the above remarks concerning the action variables we return to the question of admissible angular potentials 
$U_\sigma (\varphi )$ ( in what follows, we will skip the index "$\sigma $" as the angular dynamics does not depend on the curvature )\\
As already explained, knowing the period $T_\varphi (A)$ of the angular motion in the potential $U_\sigma (\varphi )$ as a function of the "energy" A 
one can regard the expression for the period $T_\varphi (A)$ \\

\begin{equation}
\label{e47}
T_{\varphi}(A) = 2\int_{\varphi _{min}}^{\varphi _{max}}\frac{d\varphi }{\sqrt{2(A - U(\varphi ))}}  
\end{equation}
\;\\\
as the integral equation for the potential $U_\sigma (\varphi )$. Actually, it is the integral equation of Abel type and  
it is known, (see for instance \cite{b34}) that the potential  $U(\varphi )$ is implicitly given by the relations
\;\\\
\begin{equation}
\label{e48}
\varphi _{\pm }(U) = \pm \frac{1}{2}\delta \varphi (U) + G(U) 
\end{equation}
\;\\\
where $\delta \varphi(U)$ function is given  by
\;\\\
\begin{equation}
\label{e49}
\delta \varphi(U) = \frac{1}{\pi \sqrt{2}}\int_{U_{0}}^{U}\frac{T_{\varphi }(A)dA }{\sqrt{U - A}}  
\end{equation}
\;\\\
\;\\\
and G(U) denotes a single valued function of U near $U = U_{0}$ such that the potential $U(\varphi)$ as determined by eq.(\ref{e48}) is also single 
valued.  

In our case for the periods corresponding to oscillator families and given by eq.(\ref{e43}) one finds

\begin{equation}
\label{e50}
\delta \varphi(U) = \frac{1}{2q }(\arcsin{\frac{U -2\gamma -2U_{0}}{U + \gamma }} + \frac{\pi }{2}) 
\end{equation}

while for the periods of  "Kepler" families given by eq.(\ref{e44}) one has
\begin{equation}
\label{e51} 
\delta \varphi(U) = -\frac{1}{q}(\arcsin{\frac{U - U_{0} - \sqrt{(U_{0} + B)^2 - F}}{\sqrt{(U  + B)^2 - F}}} + \frac{\pi }{2}) 
\end{equation}

So, we have shown that for any 2D space of constant curvature k  i.e. 2D sphere with $k > 0$, 2D Euclidean plane with $k = 0$ and 2D hyperbolic plane with $k < 0$ there are two families of superintegrable systems with Hamiltonians separable in the geodesic polar coordinates. These are the oscillator family (in the zero curvature case it contains isotropic harmonic oscillator) and Kepler family ( it includes the Kepler system in the $k = 0$ case). These families are defined by the radial and angular potentials. The former ones depend on the space curvature and are uniquely determined. In the case of oscillator families they are given by equations (\ref{e36}), (\ref{e38}) and (\ref{e40})  for $k < 0$, $ k = 0$ and $k > 0,$ respectively. For the Kepler families they are given by equations (\ref{e37}), (\ref{e39}) and (\ref{e41}) for $k < 0$, $ k = 0$ and $k > 0,$ respectively.\\
 On the other hand, the angular potentials corresponding to the both families do not depend on the space curvature and, contrary to the radial ones,  are determined implicitly up to a function.    
In the case of oscillator families they are given by eqs.(\ref{e48}) and (\ref{e50}) i.e. \\

\begin{equation}
\label{e52} 
 \varphi_{\pm}(U) = \frac{\pm 1}{2q }(\arcsin{\frac{U -2\gamma -2U_{0}}{U + \gamma }} + \frac{\pi }{2})  + G(U) .
\end{equation} 
 \;\\\
 For the Kepler families the potentials are given by eqs. (\ref{e48}) and (\ref{e51}) i.e. 
 
\begin{equation}
\label{e53} 
\varphi_{\pm}(U) = \mp \frac{1}{q}(\arcsin{\frac{U - U_{0} - \sqrt{(U_{0} + B)^2 - F}}{\sqrt{(U  + B)^2 - F}}} + \frac{\pi }{2})  + G(U)
\end{equation}
In other words, the angular dynamics of our superintegrable systems defined on the spaces of constant curvature is, in fact, curvature independent. 
\\

 In particular, putting $\gamma  = 0$, $q=n$, $n = 1,2,3...$ in equations defining oscillator families and taking 
 
\begin{equation}
\label{e54}
 G(U) = \frac{1}{2n}\arccos {\frac{n(\sqrt{\alpha}  - \sqrt{\beta })}{\sqrt{U}}} 
\end{equation}

give the following radial potentials

\begin{equation}
\label{e55}
\begin{split}
 V_-(r) = \frac{\delta }{\mid k\mid }\tanh^2({\sqrt{-k}r}) \qquad for\;\; k < 0\\
 V_0(r) = \delta r^2 \qquad for\;\; k = 0\\
 V_+(r) = \frac{\delta }{k}\tan^2{(\sqrt{k}r)}  \qquad for\;\; k > 0\\
 \end{split}
 \end{equation}
 \;\\\
 and $P\ddot oschl-Teller$ potential describing the angular dynamic
 
 \begin{equation}
 \label{e56}
 U(\varphi ) = \frac{n^2\alpha }{\cos^2{n\varphi} } + \frac{n^2\beta  }{\sin^2{n\varphi} } 
\end{equation}
\;\\\
(Note that $V_-$ and $V_+$ are pseudo-spherical and spherical Higgs potentials, respectively.)\\
In other words one deals with Tremblay, Turbiner and Winternitz (TTW) model for $k = 0$ and its generalizations on the curved spaces considered in \cite{b24}.

In a similar way, taking F = 0 = B in the potentials corresponding to the Kepler families and choosing G(U) function as above (see eq.(\ref{e54}))
one gets superintegrable models with the following radial potentials

\begin{equation}
\label{e57}
\begin{split}
 V_-(r) = \frac{\sqrt{\mid k\mid} \sqrt{D}}{\tanh{(\sqrt{-k}r)}} \qquad for \qquad k < 0\\
 V_0(r) = \frac{\sqrt{D}}{r} \qquad for\qquad k = 0\\
 V_+(r) = \frac{\sqrt{k}\sqrt{D}}{\tan{(\sqrt{k}r)}}  \qquad for\qquad  k > 0\\
 \end{split}
 \end{equation}
 \;\\\
and $P\ddot oschl-Teller$ potential as the angular one. (Here,  $V_-$ and $V_+$ are pseudo-spherical and spherical Schrodinger-Coulomb potentials, respectively). This is nothing but Post-Winternitz model for $k = 0$ and its generalizations  on the curved spaces considered in  \cite{b24}. Obviously, other choices are possible. In particular, $G(U) = \varphi _{0} = const$
 
\section{Central potentials case}
\label{s4}

We have been mainly interested in the classification and construction of non-central potentials generating superintegrable dynamics on 2D spaces of 
constant curvature. Nevertheless, our approach applies to central potentials as well leading to the Bertrand theorem in the case of Euclidean plane 
and its generalizations on curved spaces \cite{b26} - \cite{b31}. Indeed, for central potentials one has $U_\sigma (\varphi ) = 0$ and $\Delta \varphi \equiv \varphi_{max} - 
\varphi_{min} = \pi $. Hence, the angular action variables read (see eq. (\ref{e8a})) $J_{\sigma \varphi }(A) = \frac{1}{\pi }\sqrt{2A }\Delta \varphi$. 
On the other hand taking into account the explicit form of angular action variables (see eqs. (\ref{e45}) and (\ref{e46}) one arrives at the following relations

\begin{equation}
\label{e55}
 \frac{1}{\pi }\sqrt{2A }\Delta \varphi = 
\begin{cases}
 \frac{1}{q\sqrt{2}}\sqrt{A + \gamma }  \; \;\;\;\;\;\;\;\;\;\;\;\;\;\;\;\;\;\;\;\;\;\;\;\;\;\;\;\;\;\;\;\;\;\;\;\;\;\; for\; oscillator\; families\\
 \;\ \\
 \frac{1}{q\sqrt{2}}(\sqrt{A+B + \sqrt{F}} + \sqrt{A+B - \sqrt{F}})\;  for\; Kepler\; families \\
\end{cases}
\end{equation}
 
which holds provided $\gamma  = 0$, $q = \frac{1}{2}$ for potentials corresponding to oscillator families and $B = 0 =  F$, $q = 1$ for the ones of
the Kepler families. So, it follows from eqs. (\ref{e36}) - (\ref{e41}) that one has the isotropic harmonic oscillator $V_0(r) = \delta r^2$  and Coulomb potential $V_0(r) = \frac{\sqrt{D}}{r}$ in the Euclidean plane case, which is nothing but the Bertrand result. \\
For the curved spaces one obtains spherical $(k > 0)$ and pseudospherical $(k < 0)$ Higgs oscillator  (with $q = \frac{1}{2}$ ) belonging to the relevant oscillator families and given respectively, by potentials $ V_+(r) = \frac{\delta }{ k }\tan^2{(\sqrt{k}r)}$  and $V_-(r) = \frac{\delta }{\mid k\mid }\tanh^2({\sqrt{-k} r})$.\\
The relevant Kepler families are then represented by the spherical $(k > 0)$ and pseudospherical $(k < 0)$ Schrodinger-Coulomb potentials given respectively, by  $ V_+(r) = \frac{\sqrt{k}\sqrt{D}}{\tan{(\sqrt{k}r)}}$ and $V_-(r) = \frac{\sqrt{\mid k\mid} \sqrt{D}}{\tanh{(\sqrt{-k}r)}}$

\section{Summary}
\label{s5}

The paper deals with the construction and classification of 2D superintegrable systems ( i.e. systems admitting in addition to two global integrals of motion guaranteeing the Liouville integrability, the third global and independent one) defined on 2D spaces of constant curvature and separable in the so called geodesic polar coordinates. The method we propose is applicable to any value of curvature including the case of Euclidean plane, the 2-sphere and the hyperbolic plane. The mathematics we use is essentially "physical": in particular we refer to the very elegant technique of action-angle variables, perturbation theory and most of our mathematical formulas have a clear physical meaning. The main result can be considered as a kind of generalization of the Bertrand's theorem on 2D spaces of constant curvature.
\par We start with the general Hamiltonian separable in the polar geodesic coordinates and assume from the very beginning that it is Liouville integrable with Hamiltonian itself and generalized momentum (see eq.(\ref{e7})) providing the relevant involutive integrals. Then we refer to the necessary and sufficient condition on an integrable system to be superintegrable. In terms of the action-angle variables it says that Hamiltonian of superintegrable system has to depend on the action variables through their linear combination with integer coefficients. This condition, when applied to 2D integrable and separable (in geodesic polar coordinates) systems, results in eq.(\ref{e16}) which, in turn, implies that the derivative of radial action with respect to the A parameter (a value of the generalized momentum) does not depend on the energy. After taking into account the form of action variables corresponding to our systems and making the appropriate change of variables (see eq.(\ref{e17})) the key implication of the superintegrability condition may be written in the form of eq.(\ref{e23}). It is, up to integer factor $q$, the equality of the radial and angular periods of motions with the radial one being energy independent and hence corresponding to the dynamics in isochronous effective potentials given by eq.(\ref{e18}). This simple condition plays the crucial role in our construction. Indeed, it follows that, in order to deal with superintegrable systems, one has to find potentials $\tilde{V_\sigma (\rho ) } $ leading to isochronous effective radial potentials $W_{eff}(\rho )$. This has been done in the section four where the relevant differential equation for $\tilde{V_\sigma (\rho ) } $  has been found (see eq.(\ref{e33})) and solved. In this way, after returning to the original radial variables relevant for the given curvature sign one arrives at the radial potentials  $V_\sigma (r ) $ (see eqs.(\ref{e36} - \ref{e41}) leading to superintegrability. In the next intermediate step, knowing the explicit form of the radial potentials and referring again to eq.(\ref{e24}) or eq.(\ref{e16}), the periods of angular motions  have been calculated. Then, regarding  the expression  for a period of one-dimensional motion in a potential as the integral equation ( of the Abel type ) with the potential entering as unknown function we have determined the angular potentials corresponding to the angular periods found in the intermediate step. \\
The radial and angular potentials obtained in this way allow to conclude that for 2D spaces of constant curvature $k$: Euclidean plane ($k = 0$), the 2-sphere ($k>0$) and the hyperbolic plane ($k<0$) there exist, respectively, two families of superintegrable systems with Hamiltonians separable in geodesic polar coordinates. In the zero curvature case these families are those which were found, in different way, many years ago by Onofri and Pauri \cite{b25}. One of these families contains, in particular, isotropic oscillator and hence was called oscillator family while the other one includes the Kepler system and was named the Kepler family. We have followed this nomenclature.\\
The radial potentials we have identified depend on the curvature and are uniquely determined. In the case of the oscillator families they are given by eq. (\ref{e36}) for $k<0$,  eq. (\ref{e38}) for $k=0$ and eq. (\ref{e40}) for $k > 0$. The radial potentials of the Kepler families are given by  eq. (\ref{e37}) for $k<0$,  eq. (\ref{e39}) for $k=0$ and eq. (\ref{e41}) for $k > 0$. The angular potentials for both families, contrary to the radial ones, do not depend on the space curvature and are determined implicitly  up to an, in principle arbitrary, function. For the oscillator families they are given by eq.(\ref{e52}) while for the Kepler families by eq. (\ref{e53}).\\
In particular, we have demonstrated that choosing in the appropriate way the constants entering our potentials (i.e. $\gamma = 0$ and $q = n$ in oscillator families case and $ F = 0 = B $ in the Kepler ones ) and taking the function G in the form given by eq.(\ref{e55}) one arrives at TTW and PW models, respectively (if $k = 0$) and its spherical ($k > 0$) and pseudo-spherical ($k < 0$) generalizations. This shows that these models belong to our general families.\\
Taking into account the explicit form of radial action variables corresponding to the radial potentials defining our superintegrable systems and using
the superintegrability condition we have also presented the explicit form of angular action variables. Having these actions allows one to verify in the independent way the form of periods of radial and angular motions respectively, as well as to show that the q parameter ( the ratio of these periods ) controls the range of angle variables.\\
Finally, we also demonstrated that our approach, when applied to the central potentials case, reproduces the Bertrand theorem for Euclidean plane and its generalizations for 2D configuration spaces of constant curvature.
\;\\\
\;\\\
\;\\\

{\bf Acknowledgments}
The authors would like to thank Joanna Gonera and Piotr Kosi\'nski for helpful discussions and reading the manuscript. Useful remarks of Krzysztof Andrzejewski and Pawel Maslanka are kindly acknowledged.


\begin{thebibliography}{99}

\bibitem{b1}Arnold  V. I. 1978 Mathematical Methods of Classical Mechanics (New York: Springer-
Verlag)\\
Goldstein H, Poole C P and Safko J L 2001 Classical Mechanics (Addison Wesley) 3rd Edition\\
Perelomov A., M., ( 1990) Integrable Systems of Classical Mechanics and Lie Algebras, Birkhauser,\\
Babelon O., Bernard D., Talon M. (2007), Introduction to Classical Integrable Systems (Cambridge Monographs on Mathematical Physics)
\bibitem{b2}Runge C.(1923)  Vector Analysis (Methuen and Company  Ltd.)\\
Lenz W. (1924) Z. Physik 24, 197\\
\bibitem{b2a}Fradkin D. M. (1965) Am. J. Phys. 33, 207
\bibitem{b3}Pauli W. (1926) Z. Physik 36, 336\\
Fock V. (1935)  Z. Physik  98, 145\\
Bargmann V. (1936) Z. Physik 99, 576
\bibitem{b4}Jauch J. M., Hill E.L. (1940) Phys. Rev. 57, 641
\bibitem{b5}Evans N. W. 1990 Superintegrability in classical mechanics Phys. Rev. A. 41, 5668\\
Tempesta P, Winternitz P, Miller W and Pogosyan G, eds.( 2004) Superintegrability
in Classical and Quantum Systems vol. 37 of CRM Proceedings and Lecture Notes
(Providence, RI: AMS)\\
Miller, Jr. W., Post S., Winternitz P. (2013) Classical and Quantum Superintegrability with
Applications, arXiv: 1309.2694
\bibitem{b6}Bertrand J (1873) Theoreme relatif au mouvement d'un point attire vers un centre fixe
C. R. Acad. Sci. 77 849 – 853
\bibitem{b7}Ranada M. F . , Santander M. (1999) Superintegrable systems on the two-dimensional sphere S2
and the hyperbolic plane H2 J. Math. Phys. 40, 5026
\bibitem{b8}Ranada M. F. , Santander M. ( 2002) On harmonic oscillators on the two-dimensional sphere S2
and the hyperbolic plane H2 J. Math. Phys. 43,  431
 \bibitem{b9}Ranada M F and Santander M (2003) On harmonic oscillators on the two-dimensional sphere S2
and the hyperbolic plane H2 II J. Math. Phys. 44,  2149
\bibitem{b9a}Galajinsky A.,Lechtenfeld O.,On two-dimensional integrable models with a cubic or quartic integral of motion,
JHEP 1309, (2013),113 
\bibitem{b10}Saksida P (2001) Integrable anharmonic oscillators on spheres and hyperbolic spaces Nonlinearity
14,  977
\bibitem{b11}Ballesteros A, Herranz F J and Musso F (2013) The anisotropic oscillator on the 2d
sphere and the hyperbolic plane Nonlinearity 26,  971
\bibitem{b12}Maciejewski A. J, Przybylska M. and Yoshida H. (2010) Necessary conditions for classical
super-integrability of a certain family of potentials in constant curvature spaces J.Phys. A 43, 382001
\bibitem{b13}Tremblay F, Turbiner A,, Winternitz P (2009) An infinite family of solvable and
integrable quantum systems on a plane J. Phys. A: Math. Theor. 42, 242001
\bibitem{b14}Tremblay F, Turbiner A. V. , Winternitz P (2010) Periodic orbits for an infinite family
of classical superintegrable systems J. Phys. A: Math. Theor. 43, 015202
\bibitem{b15}Post S.,  Winternitz P. ( 2010) An infinite family of superintegrable deformations of
the Coulomb potential J. Phys. A: Math. Theor. 42, 222001
\bibitem{b16}Quesne C. (2010 ) Superintegrability of the Tremblay-Turbiner-Winternitz quantum
Hamiltonian on a plane for odd k J. Phys. A.: Math. Theor. 43, 082001
\bibitem{b17}Kalnins E. G., Kress J. M., Jr W. M. ( 2010) Superintegrability and higher order integrals
for quantum systems J. Phys. A: Math. Theor. 43, 265205
\bibitem{b18}Kalnins E. G., Kress J. M.,  Jr W. M.( 2010) Tools for verifying classical and quantum
superintegrability SIGMA 6
\bibitem{b19}Kalnins E. G., Kress J. M., Miller Jr W.( 2010) Families of classical superintegrable
systems J. Phys. A 43, 092001
\bibitem{b20}Gonera C. (2012) On the superintegrability of TTW model , Physics Letters A 376 (2012) 2341
\bibitem{b21}Celeghini E., Kuru S., Negro J.,  del Olmo M A (2013) A unified approach to quantum
and classical ttw systems based on factorizations, Annals of Physics 332, 27
\bibitem{b22}Ranada M. F. ( 2012) A new approach to the higher order superintegrability of the
Tremblay–Turbiner–Winternitz system ,J. Phys. A: Math. Theor. 45, 465203
\bibitem{b23}Ranada M. F. (2013) Higher order superintegrability of separable potentials with a new
approach to the Post–Winternitz system, J. Phys. A 46, 125206
\bibitem{b24}Hakobyan T,, Lechtenfeld O,, Nersessian A,, Saghatelian A., and Yeghikyan V.,( 2012) Integrable generalizations of oscillator and Coulomb systems via action-angle variables,Phys. Lett. A 376, 679
\bibitem{b25}Onofri  E., Pauri M. (1978) Search for periodic Hamiltonian flows: a generalized Bertrnd's theorem, J. Math. Phys. 19, 1850
\bibitem{b26}Schr¨odinger, E. (1940): Eigenvalues and eigenfunctions. Proc. Roy. Irish Acad. Sect. A 46, 9
\bibitem{b27}Higgs P. W. (1979) Dynamical symmetries in a spherical geometry I  J. Phys. A: Math. Gen. 12,309
\bibitem{b28}Leemon H. I.( 1979) Dynamical symmetries in a spherical geometry II J. Phys. A: Math. Gen. 12,489
\bibitem{b29}Granovskii Y. I., Zhedanov A. S.,  Lutsenko I. M., (1992) Quadratic algebras and dynamics
in curved spaces. i. Oscillator, Theoret. and Math. Phys. 91, 474
\bibitem{b30}Granovskii Y. I., Zhedanov A. S.,  Lutsenko I. M., (1992) Quadratic algebras and dynamics in curved spaces. ii. The Kepler problem Theoret. and Math. Phys. 91, 604
\bibitem{b31}Shchepetilov, A.V.(2005) Comment on "Central potentials on spaces of constant curvature: The Kepler problem on the two-dimensional sphere S2 and the hyperbolic plane H2" [J.Math. Phys. 46, 052702, J. Math. Phys. 46, 114101 
\bibitem{b32}Ballesteros A., Herranz F. (2007) Universal integrals for superintegrable systems on N-dimensional spaces of constant curvature J. Phys. A: Math. Theor. 40, F51
\bibitem{b33}Ballesteros A,. Herranz F., Santander M., Sanz-Gil T. (2003) Maximal superintegrability
on n-dimensional curved spaces J. Phys. A 36 L 93
\bibitem{b34}Landau L.,Lifshitz E., (1976) Mechanics, Pergamon Press,

 

\end{thebibliography}
\end{document}